\documentclass[aps,prb,preprint,12pt]{revtex4-1}
\usepackage{tikz}
\usetikzlibrary{calc}
\usepackage{amssymb}
\begin{document}

\title{On ``the'' electric field of a uniformly accelerating charge} 
\author{David Garfinkle}
\email{garfinkl@oakland.edu}
\affiliation{Department of Physics, Oakland University, Rochester, MI 48309}

\date{\today}

\begin{abstract}
The problem of the electric field of a uniformly accelerating charge is a longstanding one that has led to several issues.  We resolve these issues using techniques from linguistics, cognitive psychology, and the mathematics of partial differential equations.
\end{abstract}

\maketitle

\section{Introduction}
Accelerated charges give rise to electromagnetic radiation.  Thus one might think that a simple textbook case to illustrate this principle would be the calculation of the electric and magnetic fields of a charge undergoing constant acceleration.  However, this would-be simple textbook case is fraught with paradoxes.\cite{Born,Schott,Bondi,Rohrlich,Atwater,Gron,Almeida}  As discussed recently in reference\cite{Griffiths} the usual expression for the electric field of a moving charge, when applied to the case of uniform acceleration, does not even satisfy Maxwell's equations and needs to be ``fixed up'' by adding extra terms.  While the extra terms are considered to resolve the problem, the authors of \cite{Griffiths} are left wondering what went wrong with the standard expression.  

We will resolve this issue using a decidedly nonstandard method.  We begin with a simple linguistic observation.  The Russian language does not contain articles corresponding to the words ``the'' and ``a'' in English.  This fact tends to trip up even those native speakers of Russian who are highly fluent in English.  A somewhat helpful rule is that ``the'' is used when there is only one of something, while ``a'' is used when referring to one of a group.  Thus, we are led to the question of whether it makes sense to speak of ``the'' electric field of a collection of charges and currents. That is, for a given collection of charges and currents, is there only one possible electric field? or is there more than one?  Recall that electric and magnetic fields $\vec E$ and $\vec B$ are described by Maxwell's equations, which are given (in mks units and in vacuum) by
\begin{eqnarray}
{\vec \nabla}\cdot{\vec E} = {\frac \rho {\epsilon _0}}
\label{M1}
\\
{\vec \nabla}\cdot{\vec B} = 0
\label{M2}
\\
{\vec \nabla} \times {\vec E} = - {\frac {\partial {\vec B}} {\partial t}}
\label{M3}
\\
{\vec \nabla} \times {\vec B} = {\mu _0} {\vec J} + {\mu _0}{\epsilon _0} {\frac {\partial {\vec E}} {\partial t}}
\label{M4}
\end{eqnarray}
Here $\rho$ is the charge density and $\vec J$ is the current density.

Not every $(\rho,{\vec J})$ can act as a source for electric and magnetic fields.  To see this, take the divergence of eqn. (\ref{M4}) and apply eqn. (\ref{M1}) to obtain
\begin{equation}
{\vec \nabla} \cdot {\vec J} + {\frac {\partial \rho} {\partial t}} = 0 \; \; \; .
\label{conserve}
\end{equation}
Eqn. (\ref{conserve}) is simply the local version of the statement that charge is conserved.  Thus, conservation of charge is a property of Maxwell's equations.  

What then might we mean by asserting that $({\vec E},{\vec B})$ is ``the'' solution of Maxwell's equations for a given $(\rho ,{\vec J})$?  We might mean a mathematical statement (something like ``the mathematically unique solution'') or we might mean a physical statement (something like ``the only physically reasonable solution'').  We will consider the mathematical version first, and the physical version later.

So given a $\rho$ and a $\vec J$, is there only one $\vec E$ and $\vec B$ satisfying Maxwell's equations?  Clearly, the answer is no.  Both $({{\vec E}_1},{{\vec B}_1})$ and $({{\vec E}_2},{{\vec B}_2})$ can be solutions with the same $(\rho,{\vec J})$ provided that 
$({{\vec E}_2}-{{\vec E}_1},{{\vec B}_2}-{{\vec B}_1})$ is a solution with zero $\rho $ and $\vec J$.
These questions of uniqueness (and existence) of solutions are within the purview of the mathematical treatment of partial differential equations (PDE).  But before we unleash the beast of PDE machinery, it will be helpful to simplify things.  One of the complications of Maxwell's equations is that they consist of four equations for two vector fields $\vec E$ and $\vec B$.  This complication is usually dealt with by introducing scalar and vector potentials that automatically satisfy two of Maxwell's equations and simplify the form of the other two.  However, these potentials have their own issues of non-uniqueness having to do with gauge invariance, so we will instead apply a different strategy as follows: applying $\partial /\partial t$ to eqn. (\ref{M4}) and using eqns. (\ref{M3}) and (\ref{M1}) we obtain
\begin{equation}
- {\mu _0}{\epsilon _0} {\frac {{\partial ^2}{\vec E}} {\partial {t^2}}} + {\nabla ^2} {\vec E} = 
{\frac 1 {\epsilon _0}} {\vec \nabla}\rho + {\mu _0} {\frac {\partial {\vec J}} {\partial t}}
\label{waveE1}
\end{equation}  
Here we have also used the identity (which holds for any vector field) 
${\vec \nabla} \times ({\vec \nabla} \times {\vec E}) = {\vec \nabla}({\vec \nabla} \cdot {\vec E}) - 
{\nabla ^2} {\vec E}$. Now introducing the constant $c = 1/{\sqrt {{\epsilon _0}{\mu _0}}}$ (the speed of light) and the D'Alembertian operator $\Box$ given by 
\begin{equation}
\Box = - {\frac 1 {c^2}} {\frac \partial {\partial {t^2}}} + {\nabla ^2}
\end{equation}
we find that eqn. (\ref{waveE1}) becomes
\begin{equation}
\Box {\vec E} = 
{\frac 1 {\epsilon _0}} {\vec \nabla}\rho + {\mu _0} {\frac {\partial {\vec J}} {\partial t}}
\label{waveE2}
\end{equation}
Similarly, applying $\partial /\partial t$ to eqn. (\ref{M3}) and using eqns. (\ref{M2}) 
and (\ref{M4}) we obtain
\begin{equation}
\Box {\vec B} = - {\mu _0} {\vec \nabla} \times {\vec J}
\label{waveB}
\end{equation}
Thus each Cartesian component of the electric field (and the magnetic field) 
satisfies an equation of the form
\begin{equation}
\Box \Phi = S
\label{wave}
\end{equation}
which is known as the wave equation (with source).  Thus the question of how charges and currents make electric fields reduces to the question of how solutions $\Phi$ of eqn. (\ref{wave}) depend on the source term $S$.  

One might worry about using eqns.(\ref{waveE2}-\ref{waveB}) as a substitute for Maxwell's equations: after all what we have shown is that given a solution of Maxwell's equations (eqns. (\ref{M1}-\ref{M4})) for $({\vec E},{\vec B})$, then $({\vec E},{\vec B})$ {\emph {also}} satisfy eqn. (\ref{waveE2}-\ref{waveB}).  We have {\emph {not}} shown that given an $({\vec E},{\vec B})$ satisfying eqns. (\ref{waveE2}-\ref{waveB}), that they also satisfy Maxwell's equations.  Nonetheless, the addition of some initial conditions does give us a result of this form.  The precise statement is the following: given $(\rho ,{\vec J})$ satisfying eqn. (\ref{conserve}), and given $({\vec E},{\vec B})$ satisfying eqns. (\ref{waveE2}-\ref{waveB}), then if eqns. (\ref{M1}-\ref{M4}) are satisfied at some initial time $t_0$, they are also satisfied at all other times.  Thus the wave equation can be substituted for Maxwell's equations, provided that appropriate initial conditions are imposed.  (See appendix A for a derivation of this result).

So what does the mathematical theory of PDE tell us about the wave equation?  It tells us that the field $\Phi$ is determined by $S$ {\emph {and}} by initial data: the values of $\Phi$ and its time derivative at some initial time.  More precisely, given $S(t,{\vec x})$, a time $t_0$ and two functions ${\Phi_0}({\vec x})$ and ${P_0}({\vec x})$, there exists a unique solution of eqn. 
(\ref{wave}) $\Phi(t,{\vec x})$ for which $\Phi({t_0},{\vec x})={\Phi_0}({\vec x}) $ and 
$(\partial \Phi/{\partial t})({t_0},{\vec x})={P_0}({\vec x})$.
 
\section{Green's functions}
\label{Green}
Theorems about existence and uniqueness of solutions of PDE are profoundly satisfying to mathematicians, and at the same time profoundly unsatisfying to physicists.  We physicists like to calculate things, so if $\Phi$ is determined by $S, \, {\Phi_0}, \, $ and $P_0$ then we would like a formula that allows us to calculate $\Phi$.  Such formulas are given in terms of Green's functions.  So we want to know what Green's function corresponds to the theorem about solutions of the wave equation. The Green's function is given in eqns. (\ref{Phi1}-\ref{Phi2}) below and illustrated in Figure 1. 
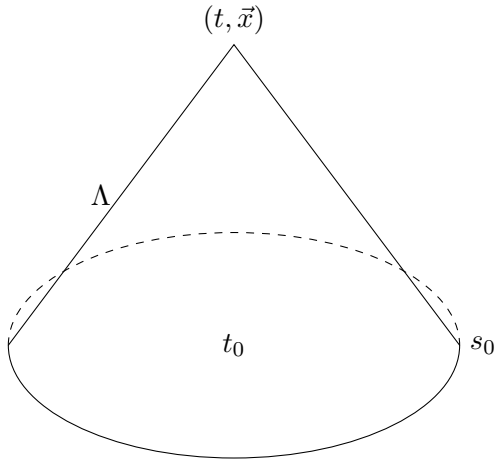
\begin{figure}
\begin{tikzpicture}
\draw (-1,0) arc (180:360:3cm and 1.5cm) -- (2,4) -- cycle;
\draw[dashed] (-1,0) arc (180:0:3cm and 1.5cm);
\node [above] at (2,4) {$(t,{\vec x})$};
\node at (2,0) {$t_0$};
\node [left] at (0.5,2) {$\Lambda$};
\node [right] at (5,0) {$s_0$};
\end{tikzpicture} 
\caption{Initial value formulation Green's function}
\label{fig:1}
\end{figure}

To start with, we write
$\Phi={\Phi_1}+{\Phi_2}$ where $\Phi_1$ is the piece of $\Phi$ determined by $S$ and $\Phi_2$ is the piece determined by $\Phi_0$ and $P_0$.  We then write separate formulas for each piece:\cite{John}
\begin{eqnarray}
{\Phi_1}(t,{\vec x}) &=& {\frac {-1} {4 \pi}} \; {\int _\Lambda} \, {d^3} {x'} \, {\frac {S({t'},{\vec {x'}})} {|{\vec x} - {\vec {x'}}|}}
\label{Phi1}
\\
{\Phi_2}(t,{\vec x}) &=& {\frac 1 {4 \pi}} \; {\int _{s0}} \, d A \, \left ( {\frac {{\Phi_0}({\vec {x'}})}
{{|{\vec x} - {\vec {x'}}|}^2}} \; + \; {\frac {{P_0}({\vec {x'}})/c + {\vec n}\cdot {\vec {\nabla}}
{\Phi_0}({\vec {x'}})} {|{\vec x} - {\vec {x'}}|}} \right )
\label{Phi2}
\end{eqnarray}
In eqn. (\ref{Phi1}) the retarded time $t'$ is given by
\begin{equation}
{t'} = t - |{\vec x} - {\vec {x'}}|/c
\label{tret}
\end{equation}
so that light emitted from the point $\vec {x'}$ at time $t'$ reaches the point $\vec x$ at time 
$t$.  In the language of relativity, the region of integration $\Lambda$ is (part of) the past light cone of the point $(t,{\vec x})$;  however (as shown in Figure 1) only that part of the light cone to the future of the $t={t_0}$ surface.  In eqn. (\ref{Phi2}) the surface $s0$ is the intersection of the past light cone of $(t,{\vec x})$ with the initial data (time ${t_0}$) hypersurface. That is, at the time of the initial data, $s0$ is a sphere on which $|{\vec x} - {\vec {x'}}|=c(t-{t_0})$.  The integration is done with respect to area 
($dA$) and the vector $\vec n$ is the unit normal to $s0$.  (Eqns. (\ref{Phi1}-\ref{Phi2}) hold when $t > {t_0}$.  Corresponding equations using advanced time instead of retarded time hold when $t < {t_0}$). 

The solution given in eqns. (\ref{Phi1}-\ref{Phi2}) differs from the usual presentation of this subject in physics textbooks.  There it is asserted that the solution of eqn. (\ref{wave}) is the retarded solution $\Phi_{\rm ret}$ given by 
\begin{equation}
{\Phi_{\rm ret}}(t,{\vec x}) = {\frac {-1} {4 \pi}} \; {\int _{\tilde \Lambda}} \, {d^3} {x'} \, {\frac {S({t'},{\vec {x'}})} {|{\vec x} - {\vec {x'}}|}}
\label{Phiret}
\end{equation}
where $\tilde \Lambda$ is the entire past light cone of the point $(t,{\vec x})$.  Despite the difference in these formulas, there is at least the possibility of agreement between the two pictures as follows: note that eqn. (\ref{Phiret}) is the limit as ${t_0} \to - \infty$ of eqn. (\ref{Phi1}).  So {\emph {if}} this limit exists, and {\emph {if}} limiting values of $\Phi_0$ and $P_0$ can be chosen to make $\Phi_2$ vanish, then $\Phi_{\rm ret}$ will at least be {\emph a} solution of the wave equation.  For the wave equation, one can simply choose $\Phi_0$ and $P_0$ to vanish, which implies the vanishing of $\Phi_2$.  However, things are not so simple for Maxwell's equations: initial data for $\vec E$ and $\vec B$ cannot be freely specified since those initial data must satisfy eqns. (\ref{M1}-\ref{M4}) at the initial time.  
One simple prescription is the following: choose initial values of $\partial {\vec E}/\partial t$ and 
$\partial {\vec B}/\partial t$ to vanish.  Then choose the initial value of $\vec E$ to be the electrostatic field that comes from the initial value of $\rho$ and choose the initial value of $\vec B$ to be the magnetostatic field that comes from the initial value of $\vec J$.  When applied to widely separated charges moving at constant velocity (i.e. the initial conditions for scattering) this prescription (see appendix B) yields a value of $\Phi_2$ that vanishes in the limit as ${t_0} \to - \infty$, and thus shows that for this sort of early behavior of the charges the retarded solution is a solution of Maxwell's equations. 

\section{The retarded ``solution''}
But even if $\Phi_{\rm ret}$ is {\emph a} solution of the wave equation, why do the textbooks call it 
{\emph {the}} solution of the wave equation?  Here the answer comes from an analog with cognitive psychology.  In \cite{Pinker}, Pinker points out that the data we receive from our senses are too incomplete and fragmentary to determine a complete picture of our surroundings.  However, survival often depends on an accurate picture of our environment, despite the paucity of sense data.  So the brain implements sophisticated algorithms, based on instinctive knowledge of how things usually are, to use sense data to produce what are essentially educated guesses about the objects in our immediate vicinity.  These guesses are presented to consciousness as perceptions, not guesses.  This is the basis for all optical illusions and stage magic.  In each case the illusionist essentially hacks into our perceptual algorithms, using tricks to induce our brains to make wrong guesses.  Similarly, as physicists we don't confine ourselves to mathematical observations about the solutions to our equations: we also apply our judgment (aka physical intuition) to decide which of those solutions are physically reasonable.  Thus when the textbooks assert that $\Phi_{\rm ret}$ is ``the'' solution of eqn. (\ref{wave}) they don't mean ``the mathematically unique solution'' but rather something more like ``the physically reasonable solution.''

But why is $\Phi_{\rm ret}$ the physically reasonable solution?  This is a deep question which is addressed well in \cite{Davies}.  In particular, note that Maxwell's equations are time reversal invariant (as is the wave equation) but that the retarded solution picks out a particular direction of time.  This is similar to the issue that Newtonian mechanics is time reversal invariant, but that the increase of entropy picks out a direction of time.   And indeed the explanation for both phenomena is essentially the same: entropy increases because the universe started out in a simple state, and the physically relevant solution of Maxwell's equations is the retarded one because the electromagnetic field started out in a simple state.  Or to put things in a more localized way: as shown by Boltzmann, entropy increase can be derived from an assumption of ``molecular chaos'' that is no pre-existing correlation between the motion of molecules. Similarly the retarded solution for the wave equation (and indeed outgoing wave motion more generally) follows from a lack of correlation in the previous state of the field.  

We are now in a position to answer the questions posed in \cite{Griffiths}.  The authors of \cite{Griffiths} want to know what goes wrong with the usual retarded solution when applied to the case of a uniformly accelerated charge.  In this section, we will consider what goes wrong in mathematical terms, and in the following section what goes wrong in physical terms.  As we have seen above, we are {\emph {not}} guaranteed that there is a retarded solution.  Rather we are guaranteed that there is a solution of eqns. (\ref{Phi1}-\ref{Phi2}) and that {\emph {if}} the appropriate ${t_0} \to - \infty$ limit of this solution exists then there is a retarded solution.  Thus the ``what goes wrong?" question of \cite{Griffiths} simply reduces to the question of what goes wrong when we try to take that ${t_0} \to - \infty$ limit.  The first thing to check is that the ${t_0} \to - \infty$ limit of eqn. (\ref{Phi1}) exists, or to put it another way that the integral in eqn. (\ref{Phiret}) converges.  In principle the question of convergence of this integral could be complicated, but things greatly simplify when we consider the electric field of a single point charge.  Because the charge moves at a speed slower than light, it can intersect the past light cone of $(t,{\vec x})$ in at most one point.  If there is one intersection point, then the entire contribution to ${\Phi_{\rm ret}}$ comes from that point.  If there is no intersection point, then ${\Phi_{\rm ret}}=0$.  The only tricky case is if there is asymptotically an intersection point in the limit as ${t'} \to - \infty$.  In that case, ${\Phi_{\rm ret}}$ is undefined.     

Thus for each $(t,{\vec x})$ our task is to figure out if the world line of the charge intersects its past light cone one time, zero times, or asymptotically as ${t'} \to - \infty$
The world line of the uniformly accelerated charge is given by ${x'}={y'}=0$ and 
\begin{equation}
{z'}={\sqrt {{b^2}+{{(c{t'})}^2}}}
\label{chargemotion}
\end{equation}
where $b$ is a constant related to the proper acceleration $a$ by $b={c^2}/a$.  The retarded time at the event where the charge world line intersects the past light cone of $(t,{\vec x})$ is given by combining eqns. (\ref{tret}) and (\ref{chargemotion}) to yield
\begin{equation}
{t'}= t \; - \; {\frac 1 c} {\sqrt {{x^2} + {y^2} + {{\left ( z - {\sqrt {{b^2}+{{(c{t'})}^2}}}\right ) }^2}}}
\label{tret2}
\end{equation}  
From eqn. (\ref{tret2}) it follows that there is no solution for $t'$ if  $ct+z < 0 $ and therefore that the retarded solution for the electric field vanishes at these points.  Furthermore, for 
$ct + z = \epsilon$ where $\epsilon$ is a small positive number, it follows from eqn. (\ref{tret2}) that 
\begin{equation}
{t'} \approx {\frac {-1} {2\epsilon}} \left ( {x^2} + {y^2} + {b^2} \right )
\label{tret3}
\end{equation}
and therefore that ${t'} \to - \infty$ as $\epsilon \to 0$.  Thus, the retarded solution exists, except at points where $ct+z=0$, and it goes discontinuously to zero there, a phenomenon described by the authors of \cite{Griffiths} as ``electric field lines ending in mid-air.'' (Though note that two figures of \cite{Griffiths} used to illustrate the properties of the electric field contain errors which are corrected in \cite{Griffiths2}). 

To understand the calculation of the previous paragraph, it is helpful to reverse the direction of time and to think of the past light cone of the point $(t,{\vec x})$ as an outgoing spherical pulse of light starting at $(t,{\vec x})$.  The position ${\vec x}'$ and time $t'$ are then the place and time at which this light pulse catches up with the moving charge.  The fact that there are events 
$(t,{\vec x})$ for which there is no interesection of the past light cone and the world line of the charge means that given a sufficiently large head start, the accelerated charge has the ability to outrun a light ray!   

Thus the retarded solution exists almost everywhere, but fails to exist at points where $ct+z=0$.  However, a function that exists almost everywhere can (provided that it is locally integrable) be regarded as a distribution.  This distribution essentially amounts to the assumption that the electric field (and the magnetic field) contains no extra delta function pieces where the original electric field was undefined ({\it i.e.} pieces proportional to $\delta (ct+z)$). We can thus ask whether the $({\vec E},{\vec B})$ defined in this way satisfy Maxwell's equations in a distributional sense.  It turns out that they do not.\cite{Bondi}  And now it is easy to see why, because the other condition needed for the retarded solution to be a solution is the existence of limiting initial data in eqn. (\ref{Phi2}) that makes 
$\Phi _2$ vanish.  However, as shown in appendix B, $\Phi_2$ is determined by those charges that at time $t_0$ are outside the past light cone of $(t,{\vec x})$.  Thus the ability of the accelerated charge to outrun a light ray allows it to remain outside the past light cone even as ${t_0} \to - \infty$.  Thus, the limit as ${t_0} \to - \infty$ of $\Phi_2$ cannot be zero in this case, and the retarded ``solution'' is not a solution.

\section{a simple solution and a not-so-simple solution}

Despite the difficulties with the retarded solution, it is simple to come up with a solution of Maxwell's equations for a uniformly accelerating charge.  Recall that at time zero, the charge is momentarily at rest at the point $(0,0,b)$.  Initial data consistent with this is that at time zero, 
$\partial {\vec E}/\partial t =0 $ and $\vec E$ is the static electric field of a point charge at  
$(0,0,b)$.  Using these initial data in eqn. (\ref{Phi2}) and the accelerating charge in eqn. (\ref{Phi1}) gives a solution of Maxwell's equations that is well behaved everywhere, without discontinuities.  Note that these initial data are also the ones that would come from the retarded solution due to a charge that remains at rest at $(0,0,b)$.  It then follows from uniqueness of solutions that the simple solution can be characterized as follows: for all $t>0$ the simple solution is the retarded solution of a charge that is at rest at $(0,0,b)$ for all ${t'}<0$ and undergoes uniform acceleration for all ${t'}>0$.  Correspondingly, for all $t<0$ the simple solution is the advanced solution due to a charge that decelerates uniformly for all ${t'}<0$ coming to rest at ${t'}=0$ and remaining at rest for all subsequent times.  Note that despite this characterization, the simple solution is a solution of the original problem: it is a solution of Maxwell's equations whose $(\rho ,{\vec J})$ is that of the uniformly accelerating charge.   

But this simple solution is not the usual answer given to this problem.  Instead the almost everywhere defined retarded fields treated in the previous section can be turned into a solution by adding extra delta function terms.\cite{Bondi,Boulware,Cross}  As noted by Boulware\cite{Boulware}, the symmetries of the problem lead to a simple ansatz for the delta function terms, and a straightforward calculation then verifies that the almost everywhere defined retarded solution plus the delta function terms is a distributional solution of Maxwell's equations.  Boulware also notes that the full solution can be obtained by the following limiting procedure:  
(1) pick some time $t_i$. (2) Consider a charge that undergoes uniform acceleration for $t>{t_i}$ and has constant velocity for $t<{t_i}$. (3) Find the retarded solution of this charge. (4) take the limit as ${t_i} \to - \infty$.  The solution of \cite{Boulware} had been found previously by Bondi and Gold\cite{Bondi} using a different limiting procedure.  This solution has also been found by Cross\cite{Cross} using the uniformly accelerated source, but performing a formal manipulation of the product of the Green's function and source involving an exchange of the order of limits. 

From our point of view, the extra delta function terms are not extra at all.  Rather they are required by the fact that the limiting solution of eqn. (\ref{Phi2}) cannot be zero.  These extra terms are rather singular.  In particular, a discontinuous electric field is turned into a solution by adding extra terms that are delta functions.

What physical interpretation should be given to the simple solution and the non-so-simple solution?  The usual physical interpretation in electromagnetism is that for a given $(\rho ,{\vec J})$ the retarded solution is ``the'' solution produced by those charges.  Any other solution is then regarded as ``the'' solution plus some extraneous, external, source-free radiation that should not be regarded as produced by $(\rho ,{\vec J})$.  Similarly, one could regard the not-so-simple solution of\cite{Bondi,Cross} as ``the'' solution of a uniformly accelerating charge.  One would then consider the simple solution as being the not-so-simple solution plus some extraneous, external radiation (which, however, has the odd property of turning the not-so-simple solution into something much smoother and much more simple).

Nonetheless, I don't think the interpretation of the previous paragraph is appropriate for the following reasons: our justification for preferring the retarded solution is that it gives the physically reasonable electric and magnetic fields.  But if the source $(\rho ,{\vec J})$ is physically {\emph {unreasonable}}, then there will be no physically reasonable electric and magnetic fields, no matter what choice we make, and thus no justification for preferring the retarded solution (suitably fixed up) over any other.  Is the uniformly accelerating charge physically unreasonable?  Certainly it has some physically unreasonable properties: it starts out in the infinite past with infinite kinetic energy.  Then it gradually comes to rest, doing an infinite amount of work in the process, after which an infinite amount of work is done on it to bring it to infinite kinetic energy in the infinite future.  Furthermore, these unphysical properties are intimately connected with the bizarre properties of the retarded ``solution'' and the need to add extra terms to it.  Recall that those properties depended on the ability of the charge to outrun a light ray.  But this ability is not possible with finite kinetic energy.  
My position is certainly {\emph {not}} that the simple solution should be regarded as ``the'' solution for a uniformly accelerating charge.  Instead I claim that the uniformly accelerating charge is physically unreasonable.  Therefore, while there are many mathematically possible solutions for the electric field of this charge, there are no physically reasonable ones.  

We can thus think of the problem of the field of a uniformly accelerated charge as being the physics equivalent of an optical illusion.  Our visual perception algorithms are primed to interpret visual data in a way that is compatible with the way the world works.  Some optical illusions induce us to make wrong interpretations, others induce us to make interpretations that are physically impossible, and still others lend themselves to two different interpretations that our mind can switch between.  Similarly, when faced with the problem of the uniformly accelerated charge, our physical intuition primes us to expect the retarded field, because that is the one that we expect to be physically reasonable.  However, the uniformly accelerated charge is not itself a physically reasonable situation, and so there is no physically reasonable electric field to be associated with it.  Rather, there are many possible electric fields, all of them physically unreasonable, and no fact of the matter at all about which one is the ``right'' one.

\section{Conclusions}

The question of ``the'' electric field of a uniformly accelerated charge is an excellent illustration of the pitfalls of idealizations.  Radiation comes from accelerated charges, and a uniformly accelerated charge is a simple idealization for the more complicated (but more realistic) case of a non-uniformly accelerated charge.  But because of infinite kinetic energy, the uniformly accelerated charge is unphysical and can lead us to ask unphysical questions and give unphysical answers.  It is somewhat ironic that the work of Bondi and Gold\cite{Bondi} was intended to clarify the issue of gravitational waves, because the history of gravitational waves is complicated by another unphysical idealization: plane waves.  We use plane waves all the time as an idealization, both as a simplification and because far from their sources realistic waves locally resemble plane waves.  But plane waves are unphysical because their infinite extent involves infinite energy.  This doesn't cause problems in electromagnetism because Maxwell's equations are linear and through superposition and Fourier transforms, realistic waves with finite energy can be written as a superposition of plane waves.  However, general relativity is non-linear, and so its plane waves cannot be superposed.  Progress on gravitational waves had to wait until a more suitable idealization (asymptotic flatness, also due to Bondi\cite{Bondi2}) was developed. 

\section{Acknowledgements}

This work was partially supported by
NSF grants PHY-1505565 and PHY-1806209 to Oakland University.

\appendix
\section{equivalence of Maxwell's equations and the wave equation}

In this appendix we show that Maxwell's equations are a consequence of (i) the wave equation for electric and magnetic field, (ii) conservation of charge, and (iii) the imposition of Maxwell's equation as an initial condition.  More precisely, we assume that $\vec E$ and $\vec B$ satisfy
\begin{eqnarray}
\Box {\vec E} &=& 
{\frac 1 {\epsilon _0}} {\vec \nabla}\rho + {\mu _0} {\frac {\partial {\vec J}} {\partial t}} \; \; \; ,
\label{waveEa}
\\
\Box {\vec B} &=& - {\mu _0} {\vec \nabla} \times {\vec J} \; \; \; .
\label{waveBa}
\end{eqnarray}
We further assume that $(\rho ,{\vec J})$ satisfy conservation of charge
\begin{equation}
{\vec \nabla} \cdot {\vec J} + {\frac {\partial \rho} {\partial t}} = 0 \; \; \; .
\label{conservea}
\end{equation}
However we do {\emph {not}} assume that Maxwell's equations are satisfied.  Instead we define quantities ${M_1}, \, {M_2}, \, {{\vec M}_3}$ and ${\vec M}_4$ given by 
\begin{eqnarray}
{M_1} \equiv {\vec \nabla}\cdot{\vec E} - {\frac \rho {\epsilon _0}}
\label{M1def}
\\
{M_2} \equiv {\vec \nabla}\cdot{\vec B} 
\label{M2def}
\\
{{\vec M}_3} \equiv {\vec \nabla} \times {\vec E} + {\frac {\partial {\vec B}} {\partial t}}
\label{M3def}
\\
{{\vec M}_4} \equiv - {\vec \nabla} \times {\vec B} + {\mu _0} {\vec J} + {\mu _0}{\epsilon _0} {\frac {\partial {\vec E}} {\partial t}}
\label{M4def}
\end{eqnarray} 
Comparison with eqns. (\ref{M1}-\ref{M4}) shows that Maxwell's equations are equivalent to the vanishing of the quantities $({M_1},{M_2},{{\vec M}_3},{{\vec M}_4})$.  Now take the time derivative of these quantities using eqns. (\ref{waveEa}), (\ref{waveBa}), and (\ref{conservea}) 
(and $c=1/\sqrt{{\epsilon _0}{\mu _0}}$) to obtain
\begin{eqnarray}
{\frac {\partial {M_1}} {\partial t}} = {c^2} {\vec \nabla} \cdot {{\vec M}_4} \; \; \; ,
\label{dtM1}
\\
{\frac {\partial {M_2}} {\partial t}} = {\vec \nabla} \cdot {{\vec M}_3} \; \; \; ,
\label{dtM2}
\\{\frac {\partial {{\vec M}_3}} {\partial t}} = {c^2} \left ( {\vec \nabla} \times {{\vec M}_4} + {\vec \nabla} {M_2} \right ) \; \; \; ,
\label{dtM3}
\\{\frac {\partial {{\vec M}_4}} {\partial t}} = {\vec \nabla} {M_1} - {\vec \nabla} \times {{\vec M}_3} \; \; \; .
\label{dtM4}
\end{eqnarray}
Eqns. (\ref{dtM1}-\ref{dtM4}) are a system of linear first order equations with constant coefficients, and it follows that if at some time $t_0$ the quantities $({M_1},{M_2},{{\vec M}_3},{{\vec M}_4})$ all vanish, then they vanish for all other times too.  Thus, provided that Maxwell's equations are satisfied at time $t_0$, they are satisfied at all times.

\section{properties of $\Phi_2$}

To satisfy Maxwell's equations at the initial time $t_0$ we make the ansatz that 
$\partial {\vec E}/\partial t$ and $\partial {\vec B}/\partial t$ vanish at that time.  Maxwell's equations then become
\begin{eqnarray}
{\vec \nabla}\cdot{\vec E} = {\frac \rho {\epsilon _0}}
\label{M1t0}
\\
{\vec \nabla}\cdot{\vec B} = 0
\label{M2t0}
\\
{\vec \nabla} \times {\vec E} = 0
\label{M3t0}
\\
{\vec \nabla} \times {\vec B} = {\mu _0} {\vec J} 
\label{M4t0}
\end{eqnarray}
Note that eqns. (\ref{M1t0}) and (\ref{M3t0}) are just the equations of electrostatics, while eqns. (\ref{M2t0}) and (\ref{M4t0}) are the equations of magnetostatics.  Thus our solution is that at $t_0$, $\vec E$ is the electrostatic field due to $\rho$ and $\vec B$ is the magnetostatic field due to 
$\vec J$.  

In electrostatics, it is usual to introduce the electrostatic potential and obtain a Poisson equation for that potential.  However, we will instead use only the electric field.  Taking the curl of eqn. (\ref{M3t0}) and using eqn. (\ref{M1t0}) we obtain
\begin{equation}
{\nabla ^2}{\vec E} = {\frac 1 {\epsilon _0}} {\vec \nabla } \rho
\label{Poisson}
\end{equation} 
Thus each component of $\vec E$ satisfies Poisson's equation.  Now let $\Phi$ be any component of $\vec E$ and consider eqn. (\ref{Phi2}) for $\Phi_2$.  Since $\partial {\vec E}/\partial t$ vanishes
at time $t_0$, it follows that ${P_0}=0$.  Define a new coordinate ${\vec r} = {{\vec x}'} - {\vec x}$ and let $R$ be the radius of the sphere that is being intrgrated over.  Then eqn. (\ref{Phi2}) becomes
\begin{eqnarray}
{\Phi_2} &=& {\frac 1 {4 \pi}} {\int _{r=R}} \, d A \; \left ( {r^{-2}} {\Phi_0} + {r^{-1}} {\vec n} \cdot {\vec \nabla}{\Phi_0} \right ) 
\nonumber
\\
&=& {\frac 1 {4 \pi}} {\int _{r=R}} \, d A \; {\vec n} \cdot \left ( {r^{-1}} {\vec \nabla} {\Phi_0} - {\Phi_0} {\vec \nabla } ({r^{-1}}) \right ) 
\nonumber
\\
&=& {\frac {-1} {4 \pi}} {\int _{r>R}} \, {d^3} r \, \left ( {r^{-1}} \, {\nabla ^2} {\Phi_0} \right ) \; \; \; .
\label{Phi2vol}
\end{eqnarray}  
In the last line we have used Green's theorem and the fact that ${\nabla ^2}({r^{-1}})=0$ to convert the surface integral into a volume integral over the exterior of the sphere.  And we have used fall off conditions of electrostatic fields to insure that there is no additional surface integral at infinity.  If there is no charge outside the sphere, then it follows from eqn. (\ref{Poisson}) that the volume integral in eqn. (\ref{Phi2vol}) vanishes, and therefore that $\Phi_2$ vanishes.  As 
${t_0} \to - \infty$ we have that $R$ grows at the speed of light.  Thus any charge with a limitting velocity slower than light will eventually (i.e. for sufficiently large and negative $t_0$) be inside the sphere and cease to contribute to $\Phi_2$.  If all charges have this behavior, then 
${\lim_{{t_0} \to - \infty}} {\Phi_2} = 0$.

\end{document}